\documentclass[runningheads]{llncs}

\usepackage[utf8]{inputenc} 
\usepackage[T1]{fontenc}    
\usepackage[colorlinks, linkcolor=blue]{hyperref}      
\usepackage{url}            
\usepackage{booktabs}       
\usepackage{amsfonts}       
\usepackage{nicefrac}       
\usepackage{graphicx}
\usepackage{subcaption} 
\usepackage[english]{babel}
\usepackage{doi}
\usepackage{amsmath}
\usepackage{listings}
\usepackage{xcolor}
\usepackage[inkscapelatex=false]{svg}
\usepackage{fontawesome5}
\usepackage{marvosym}
\usepackage{float}

\newcommand \footnoteONLYtext[1]
{
    \begingroup
    \renewcommand\thefootnote{}\footnote{#1}
    \addtocounter{footnote}{-1}
    \endgroup

}
\begin{document}
\title{YOLO-Angio: An Algorithm for Coronary Anatomy Segmentation}
%
%
\author{Tom Liu \inst{1, 2}\textsuperscript{(\Letter)}\orcidID{0000-0003-2351-236X}\and 
Hui Lin\inst{3}\orcidID{0000-0002-6559-2751}\and 
Aggelos K. Katsaggelos\inst{3}\orcidID{0000-0003-4554-0070} \and  
Adrienne Kline\inst{1, 2}\orcidID{0000-0002-0052-0685}
}
\authorrunning{T. Liu et al.}
%
\institute{Center for Artificial Intelligence, Bluhm Cardiovascular Institute, Northwestern Medicine, Chicago IL 60611, USA \and
Department of Cardiac Surgery, Northwestern University, Chicago IL 60611, USA \and
Department of Electrical and Computer Engineering, Northwestern University, Evanston IL 60208, USA\\
\email{\textsuperscript{\Letter}tom.liu@northwestern.edu}\\
}

\maketitle              
\footnoteONLYtext{T. Liu and H. Lin—The authors contributed equally to this work.}
\begin{abstract}

Coronary angiography remains the gold standard for diagnosis of coronary artery disease, the most common cause of death worldwide. While this procedure is performed more than 2 million times annually, there remain few methods for fast and accurate automated measurement of disease and localization of coronary anatomy. Here, we present our solution to the Automatic Region-based Coronary Artery Disease diagnostics using X-ray angiography images (ARCADE) challenge held at MICCAI 2023. For the artery segmentation task, our three-stage approach combines preprocessing and feature selection by classical computer vision to enhance vessel contrast, followed by an ensemble model based on YOLOv8 to propose possible vessel candidates by generating a vessel map. A final segmentation is based on a logic-based approach to reconstruct the coronary tree in a graph-based sorting method. Our entry to the ARCADE challenge placed 3rd overall. Using the official metric for evaluation, we achieved an F1 score of 0.422 and 0.4289 on the validation and hold-out sets respectively. 

\end{abstract}
\section{Introduction}
\label{Introduction}
Coronary angiography is used to diagnose and treat coronary artery disease (CAD), a condition where atherosclerotic plaque builds up in the blood vessels of the heart \cite{CAD}. During this procedure, luminal narrowing, plaque characteristics, and real-time blood flow are observed following contrast injection. Data are recorded as time series images, which are used for both diagnosis of disease and as a guide for procedures such as stent insertion. The recorded images assist cardiologists and other healthcare professionals in assessing the severity of coronary disease and the expected success of percutaneous interventions. Although angiography is the gold standard for the assessment and management of CAD, interpretation of angiogram images has high intra- and inter-rater variability \cite{irr_syntax}. Assessments are made by qualitative examination of contrast flow through branches of the arterial tree. Thus, the objective of coronary angiograms is a key target for automation.

\noindent 
There are very few works aimed at automating the diagnosis of coronary artery disease using angiography. Early work in this area includes vessel boundary detection \cite{angionet_segmentation,gao_segmentation}, centerline extraction \cite{wang_centerline,habijan_centerline} and registration \cite{registration}. However, these methods are generally not automated and are not clinically useful since they provide little anatomic context to assist with the diagnosis of CAD. In the clinical setting, prior work has focused on using human input to selectively segmenting regions of interest and calculating the degree of stenosis, a practice known as quantitative angiography \cite{quantitative_angio}. The user will define relevant anatomic regions for analysis, the software will then provide a numerical percentage calculated that can be used to assess disease severity. Thus, there remains a need for fast automated segmentation and identification of clinically relevant anatomic locations together. This kind of application has the potential to become diagnostically and clinically useful. 

\noindent 
Current state-of-the-art segmentation that incorporates relevant anatomic features include CNN and U-Net-based approaches, which offer high discriminatory ability for vessel boundary lines \cite{SOTA_unet}. CathAI is a multiple CNN-based approach using multimodal input to estimate stenosis and identify 11 major segments and sub-segments \cite{cathai} in the coronary tree. DiscernNet \cite{china_eurointervention} uses an adversarial network with a large repository of images to accurately identify 20 of 25 major anatomic segments with high reported accuracy and discriminatory ability. However, both are notably trained on a private dataset, and no true benchmark exists in a large clinically annotated public dataset. 

\noindent 
In this paper, we report a fast segmentation of the coronary artery tree for the Automatic Region-based Coronary Artery Disease diagnostics using x-ray angiography images (ARCADE) challenge held at MICCAI 2023. The overall pipeline of our method is shown in Figure \ref{fig:overview}, it consists of three main stages: (1) in the first stage, we use various image transformations to enhance vessel and feature contrast; (2) segmentation is performed using an ensemble model trained on image features obtained in stage 1. We train separate models based on feature transformations of the original image using YOLOv8; (3) a logic-based approach is finally used to reconstruct the final coronary artery tree in a graph-based sorting method. 

\section{Methods}

\begin{figure}[ht]
    \centering
    \includegraphics[width=1\linewidth]{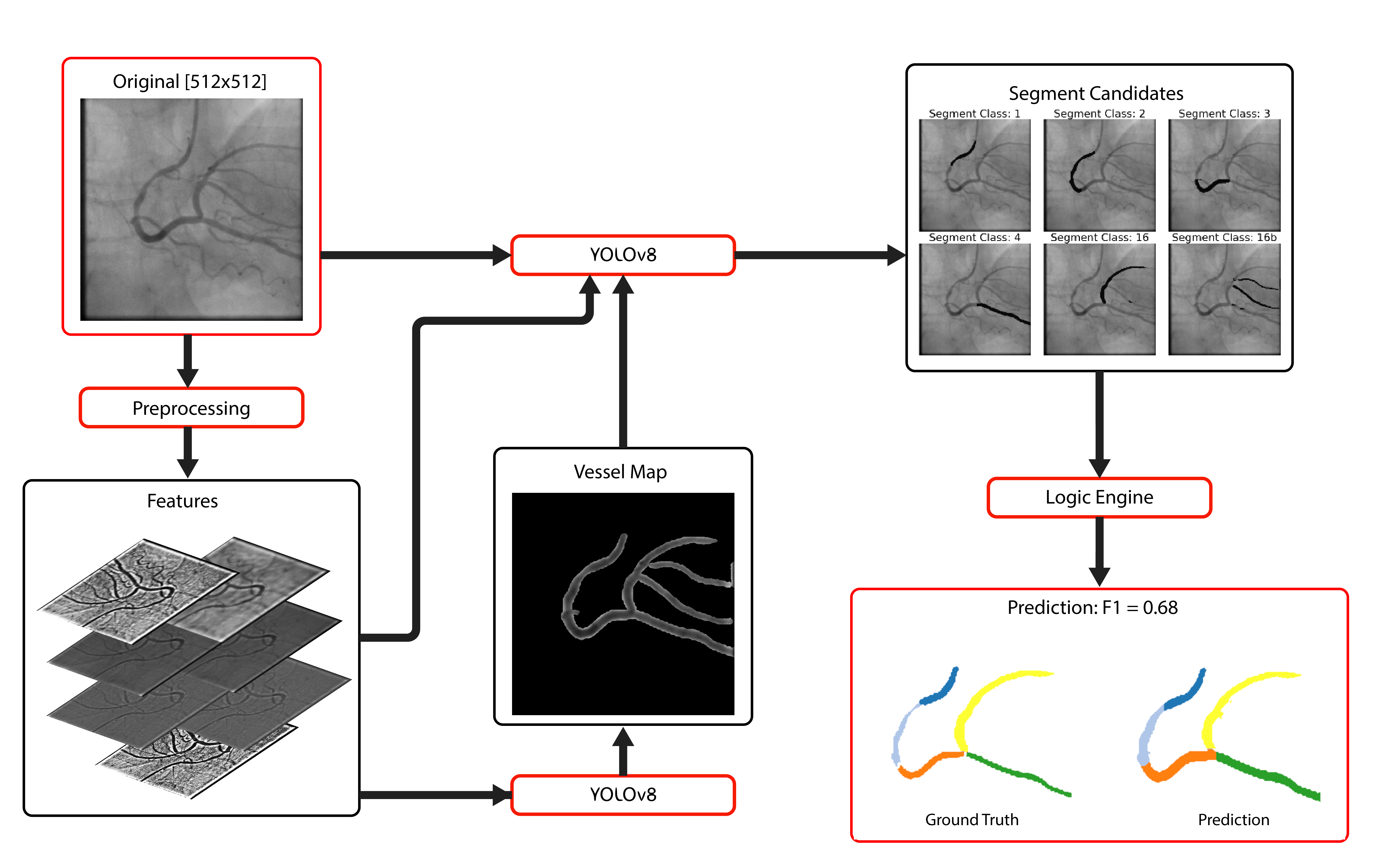}
    \caption{Overview of YOLO-Angio. Feature selection is performed to enhance vessel contrast, followed by YOLO-based segmentation using an ensemble model and a logic-based approach to construct the final coronary artery tree.}
    \label{fig:overview}
\end{figure}

We apply a 3 stage framework to perform segmentation of the coronary arterial tree, as shown in Figure \ref{fig:overview}.

\subsection{Dataset}
\label{dataset}

Our framework is trained and tested using the 2023 ARCADE Challenge dataset\footnote[1]{\href{https://arcade.grand-challenge.org/} {https://arcade.grand-challenge.org/}}  \cite{maxim_popov_2023_8378645} which contains 1000 annotated angiogram images at peak contrast for training and 200 for validation as well as 300 for testing. Input images are static 2-dimensional, 512$\times$512 pixels at various magnifications of the coronary tree. Coronary artery segments are labeled according to SYNTAX score definitions \cite{sianos_syntax}, which denote 25 coronary artery segments according to the Arterial Revascularization Therapies Study (ARTS) trial \cite{arts}. The frequency distribution of these structures in the ARCADE dataset is shown in Figure \ref{figure: class_distribution}. 

\subsection{Image Preprocessing and Feature Selection}
\label{preprocessing}

Figure \ref{figure: preprocess_example} shows contrast enhancement and feature selection of images to highlight features of the coronary arterial tree and reduce background noise. Image enhancement is important for the analysis of angiographic images due to high levels of noise and varying levels of contrast. This step can improve the identification of region boundaries as stated earlier \cite{popov2022review}. The goal of enhancement is to obtain a better vessel map for subsequent segmentation. Images were processed in the following way:

\begin{figure}[ht]
    \centering
    \includegraphics[width=0.95\linewidth]{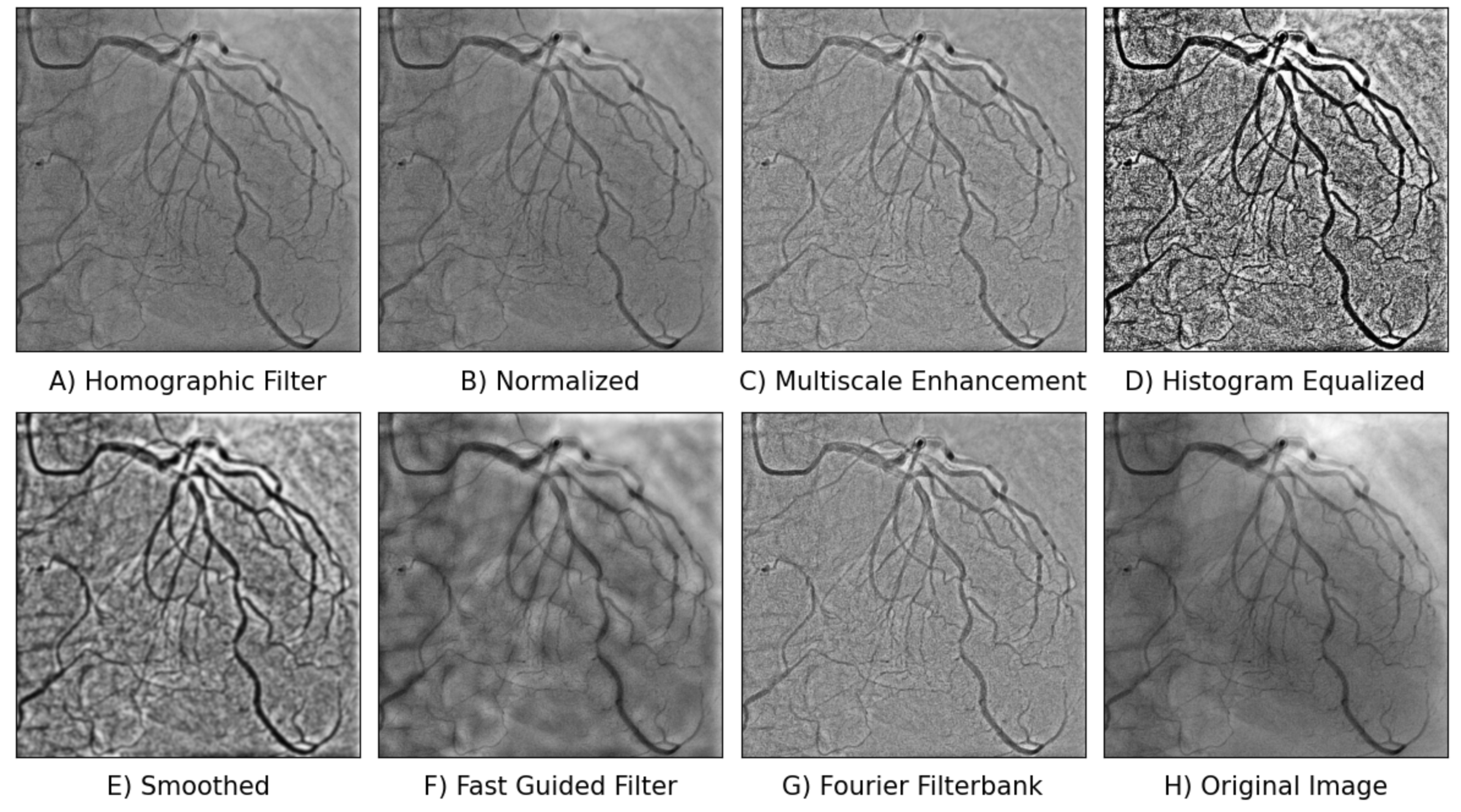}
    \caption{A. Homomorphic filtered image. B. Normalized image. C. Multiscale top hat enhancement following high pass filter using homomorphic filter. D. Histogram equalization and E. Smoothing. F. Guided filter. G. Directional enhancement with Fourier filter bank. H. Original image.}
    \label{figure: preprocess_example}
\end{figure}

\noindent \\ 
A) Homomorphic image enhancement is performed with a Butterworth high pass filter to reduce background noise and glare as described in \cite{ddfb_khan}. Briefly, in Eq. (\ref{equation: illuminance}) an input image \(I(x, y)\) can be expressed as the sum of its log-transformed illuminance \(\ln(i(x, y))\) and reflectance \(\ln(r(x, y))\), where $(x,y)$ denotes spatial coordinates. A Butterworth low pass filter of order $n$, and with cutoff frequency $D_0$ and $D(u,v)=\sqrt(u^2+v^2)$, with $(u,v)$ the discrete frequency domain coordinates, is applied (\ref{equation: butterworth}, \ref{equation: homomorphic}) in order to reduce noise.

\begin{equation}
\label{equation: illuminance}
\ln\{I(x, y)\} = \ln\{i(x, y)\} + \ln\{r(x, y)\}
\end{equation}

\begin{equation}
\label{equation: butterworth}
H(u, v) = \frac{1}{1 + \left(\frac{D(u, v)}{D_0}\right)^{2n}}
\end{equation}

\begin{equation}
\label{equation: homomorphic}
N(u,v)=H(u,v) \times {\cal{F}}\{\ln\{ I(x, y) \} \}
\end{equation}
where $\cal{F}$ denotes the 2D discrete Fourier transform.

\noindent \\ 
B) The resultant image $N(i, j)=\exp\{n(i,j)\}$, where $n(i,j)={\cal{F}}^{-1}\{N(u,v)\}$, is normalized using Eq. \ref{equation: normalization} also according to \cite{ddfb_khan}.

\begin{equation}
N(i, j) =
\begin{cases}
    M_0 + \sqrt{\frac{{\text{VAR}_0 \cdot (img - M)^2}}{{\text{VAR}}}} & \text{if } N(i, j) > M \\
    M_0 - \sqrt{\frac{{\text{VAR}_0 \cdot (img - M)^2}}{{\text{VAR}}}} & \text{otherwise}
\end{cases}
\label{equation: normalization}
\end{equation}

\noindent \\ 
C) Multiscale top hat enhancement is used to perform boundary enhancement similar to (\ref{equation: multi_scale}), where C refers to all top or black hat transforms at all scales and D refers to the max projection of all transforms.

\begin{equation}
\label{equation: multi_scale}
I_{En}=I+I_{w}-I_{b}=I+(I_{w}^{C}+I_{w}^{D})-(I_{b}^{C}+I_{b}^{D})
\end{equation}

\noindent \\ 
D) Adaptive equalization is applied to improve contrast.

\noindent \\ 
E) A Gaussian filter with a circular kernel to perform edge smoothing.

\noindent \\ 
F) Finally, a fast-guided filter \cite{he2015fast} is used on the original image to perform edge smoothing (\ref{equation: fgf}).

\begin{equation}
\label{equation: fgf}
q_i = a_kI_i + b_k, \ \ \forall i \in \omega_k
\end{equation}

\noindent \\ 
G) A decimation-free directional filter bank (\ref{equation: ddfb}) is also used to highlight directional features such as arteries in the coronary arterial tree \cite{ddfb_khan}. The image is deconstructed into directional images in the Fourier domain and a rectangularly separable high pass filter with a cutoff frequency of $\pi/16$ and a 40dB stop-band attenuation is used to divide the background noise in the image. The image with the maximum directional energy is added to the original image to enhance it. 

\begin{equation}
\label{equation: ddfb}
f_{hf}(X, Y)=max_{1\leq i\leq 8}f_{i}(X, Y)
\end{equation}

\noindent \\ 
In our current implementation, each stage of image preprocessing described is independently used to train a neural network model. 

\subsection{Network Architecture}
\label{network}

\begin{figure}[ht]
    \centering
    \includegraphics[width=0.85\linewidth]{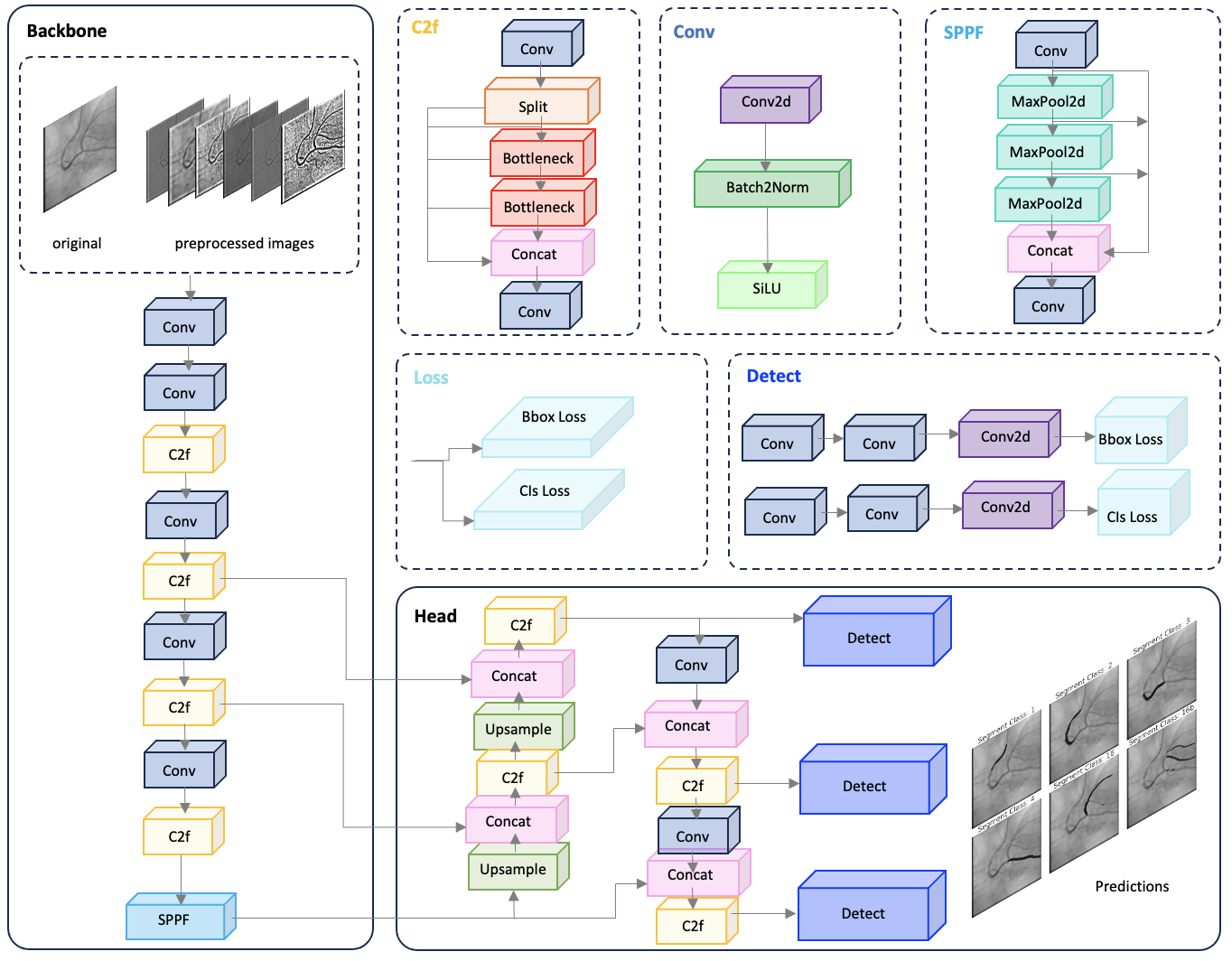}
    \caption{Architecture of YOLOv8 adapted for use in YOLO-Angio. Adapted from \cite{yolo_v5}.}
    \label{fig:yolov8}
\end{figure}

\noindent
The segmentation network involved in our framework is adapted from YOLO \cite{yolo_v5}. Briefly, YOLO consists of three key components, with a 24-layer convolutional neural network backbone pre-trained on ImageNet at a lower resolution (Fig \ref{fig:yolov8}). The object detection head uses a fully connected neural network to assist with segmentation. This architecture is well suited for high throughput fast segmentation of the coronary artery tree due to its ability to reduce false positives. In our model, we use YOLOv8, which is optimized from YOLOv5. YOLOv8 uses a combined DICE and binary cross-entropy loss.

\subsection{Logic Based Tree Segmentation}
\label{logicnet}

\begin{figure}[ht]
    \centering
    \begin{subfigure}[b]{0.6\textwidth}
        \centering
        \includegraphics[width=\textwidth]{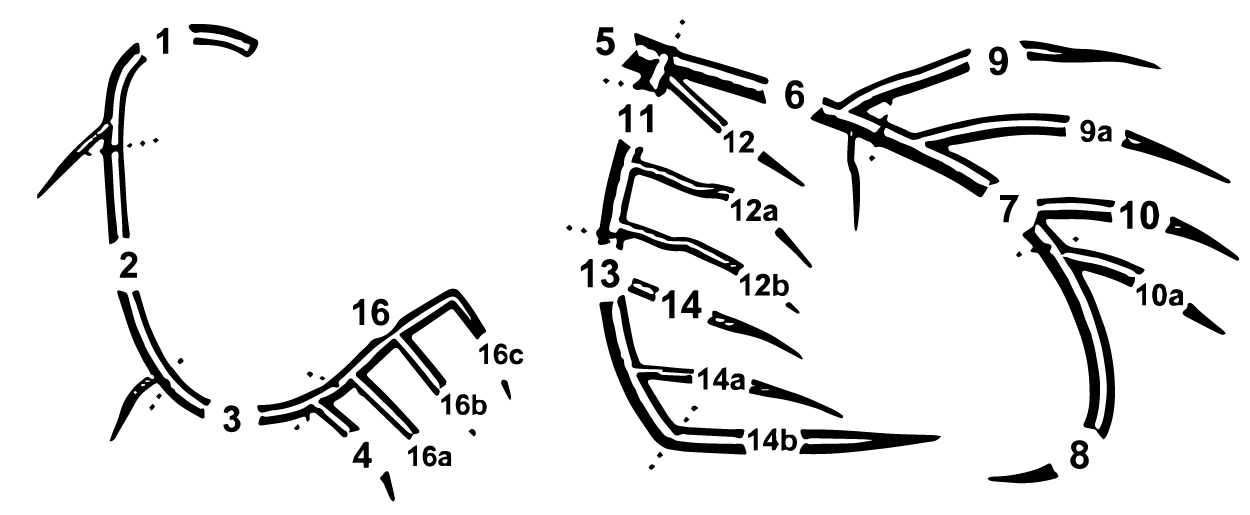}
        \caption{}
        \label{fig:left_right}
    \end{subfigure}
    \hfill
    \begin{subfigure}[b]{0.35\textwidth}
        \centering
        \includegraphics[width=\textwidth]{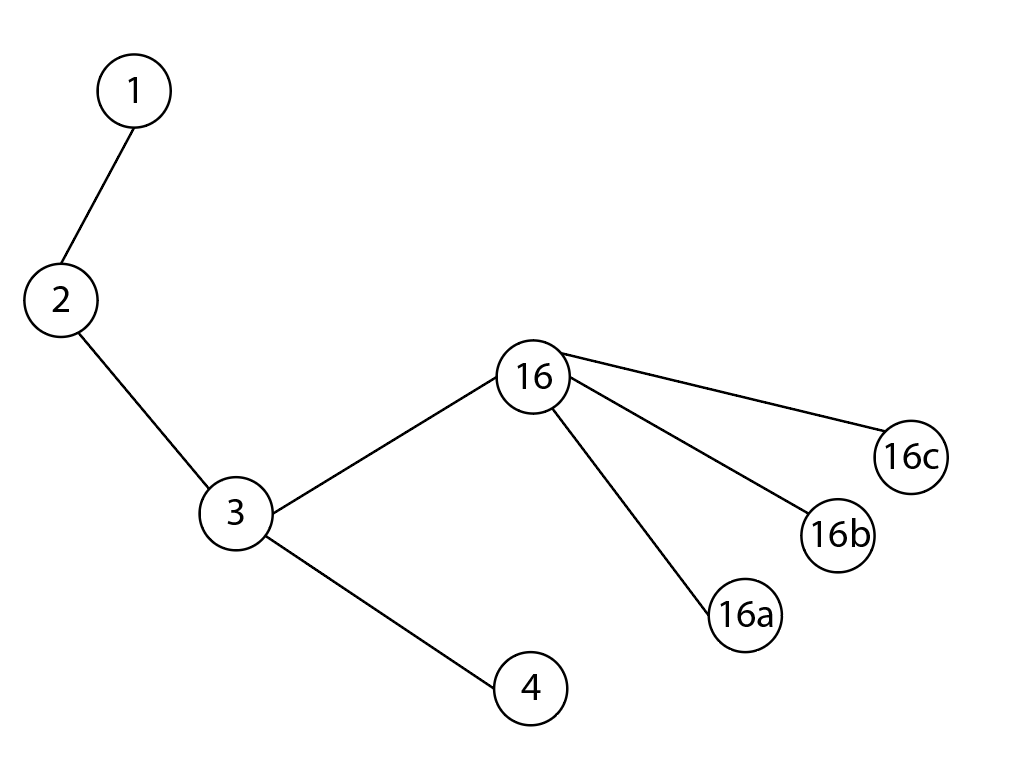}
        \caption{}
        \label{fig:CAT}
    \end{subfigure}
    \caption{A. Coronary circulation is divided into right and left circulation. B. The coronary arterial tree is represented as a graph structure with segments as vertices, using the right coronary circulation as an example.}
    \label{fig:grouped_figures}
\end{figure}

\noindent \\ 
In order to further reduce false positives, a logical-based tree segmentation is employed to confirm candidate segments. Important features of the logical-based segmentation include 1) Internal validation that predicted segments conform to either left or right heart anatomy (shown in Figure \ref{fig:left_right}) and 2) A logical graph-based construction that validates predictions from the closest segment to the aorta to the furthest sub-segment from the aorta.

\begin{figure}[ht]
    \centering
    \includegraphics[width=0.95\linewidth]{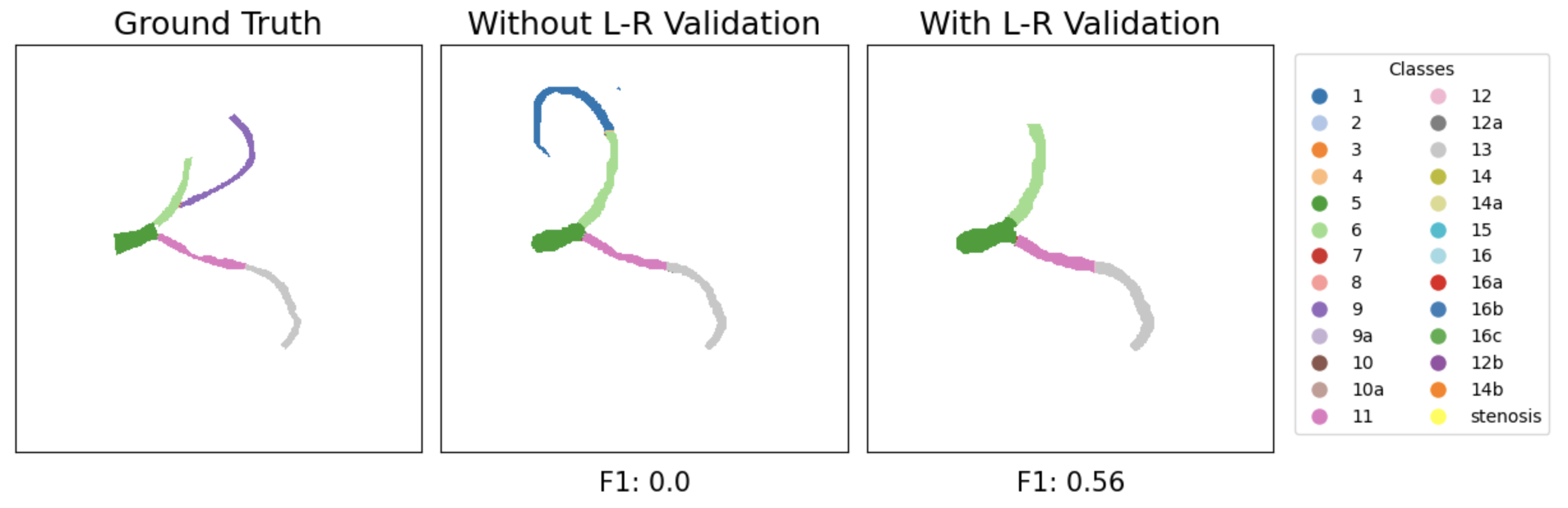}
    \caption{Removal of segment 1 (middle panel – blue) from a left anatomic structure using logical validation. The F1 score is improved due to the removal of an incorrect prediction.}
    \label{fig:LR_Logic}
\end{figure} 

\noindent \\ 
The coronary arterial tree can be further divided into a left and a right circulation (Fig \ref{fig:left_right}), which helps to reduce false positives. At the very basic level, the coronary arterial tree can be generalized as a graph object, with vertices being the demarcation between parent and child segments (Fig \ref{fig:CAT}). For example, segment 5 is the parent of segments 6, 11, and 12. This graph structure can be used to further enforce class inclusion based on the coronary tree anatomy. Similarly, in a left-sided circulation involving the left coronary arterial branch (5), it is not possible to have a 1, 2 or 3 segment from the right coronary arterial tree (Fig \ref{fig:LR_Logic}). This post-detection logic can be attached as a tail to any detection model to further improve class inclusion in segmentation. Finally, post-detection logic can be used to further sort segments that are commonly misclassified such as 9, 9a, and 10 or 16, 16a, and 16b.

\section{Experiments}
\label{experiments}

In our work, all experiments were conducted on a workstation equipped with a single NVIDIA Titan RTX GPU card with 24 GB memory. The YOLO model was trained for a multi-classification goal (26 classes) using an AdamW optimizer. For the learning rate schedule, a polynomial decay learning rate schedule was adopted, with an initial learning rate set to 0.01. The use of a polynomial decay learning rate schedule in training offered the advantage of achieving stable convergence by gradually reducing the learning rate over time, thereby enabling better fine-tuning and improved final performance. Each model in the ensemble was trained for 450 epochs, with a dropout of 0.5 to mitigate overfitting. Data augmentation included the default configuration used in YOLO documentation\footnote[2]{\href{https://docs.ultralytics.com/} {https://docs.ultralytics.com/}}. The smallest YOLO model was selected (3.4M parameters) for size and speed of inference. Training each model took approximately 1 hour.  

\subsection{Implementation details}
\label{Metrics}

Image preprocessing used the following parameters: For homographic filtering, the cutoff was $D_0$ = 12 as the distance from the origin. For image normalization, the mean, and variance were set to $M_0$ = 128 $VAR_0$ = 100, respectively. For multiscale top hat enhancement, the structuring elements were disk shapes with varying sizes from 3 to 19. For Gaussian smoothing, \(\sigma\) = 2. For the fast guided filter, r = 8 and regularization parameters \(\epsilon\) = 0.2 \(\alpha\) = 2 are used. Finally, for the directional filter -bank, a rectangularly separable high pass filter with a cutoff frequency of $\pi/16$ and a 40dB stop-band attenuation was used.

\begin{figure}[ht]
    \centering
    \includegraphics[width=0.5\linewidth]{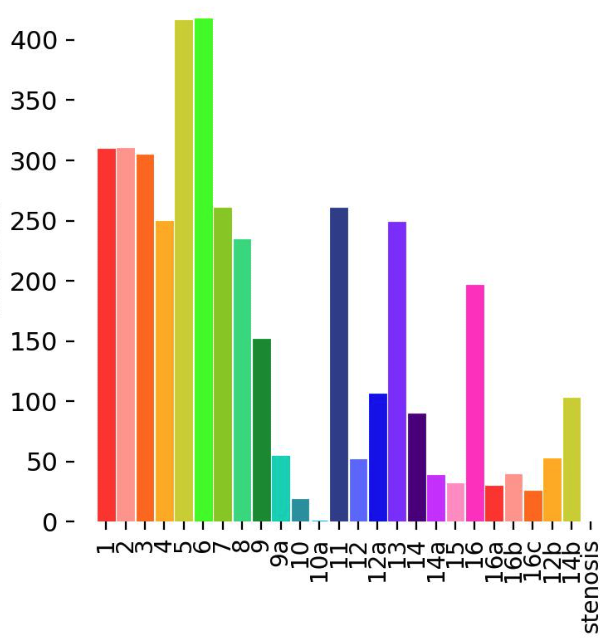}
    \caption{Distribution of Classes in Training Data. Notably, segment 10, 10a, 14a, 15, 16, 16a, 16b, 16c, and 12b were excluded due to inadequate training data}
    \label{figure: class_distribution}
\end{figure}

\noindent \\ 
The training dataset was evaluated to examine the distribution of classes. In the prediction challenge, we excluded classes with less than 25 instances in the training data (Fig \ref{figure: class_distribution}) to reduce false positives. To further reduce false positives, an area filter was applied to remove detected objects less than 450 pixels in size.

\begin{table}[ht]
\begin{center}
\caption{Algorithm Performance and Ablation Study}
\label{table: ablation study}
\begin{tabular}{c|c|c}
    \toprule
    Method & F1 & Comment \\
    \midrule
    \midrule
    Original Image & $0.271$ & YOLO model trained on original image alone \\
    Histogram Equalization & $0.309$ & Best Single Model Performance\\
    \midrule
    Ensemble w/out Logical Connector & $0.366$ & Graph-based construction is ablated. \\
    Ensemble w/ Logical Connector & \boldmath$0.411$ & \boldmath$0.4289$ in the Leaderboard. \\
    \bottomrule
\end{tabular}
\end{center}
\end{table}

\begin{figure}[ht]
    \centering
    \includegraphics[width=1\linewidth]{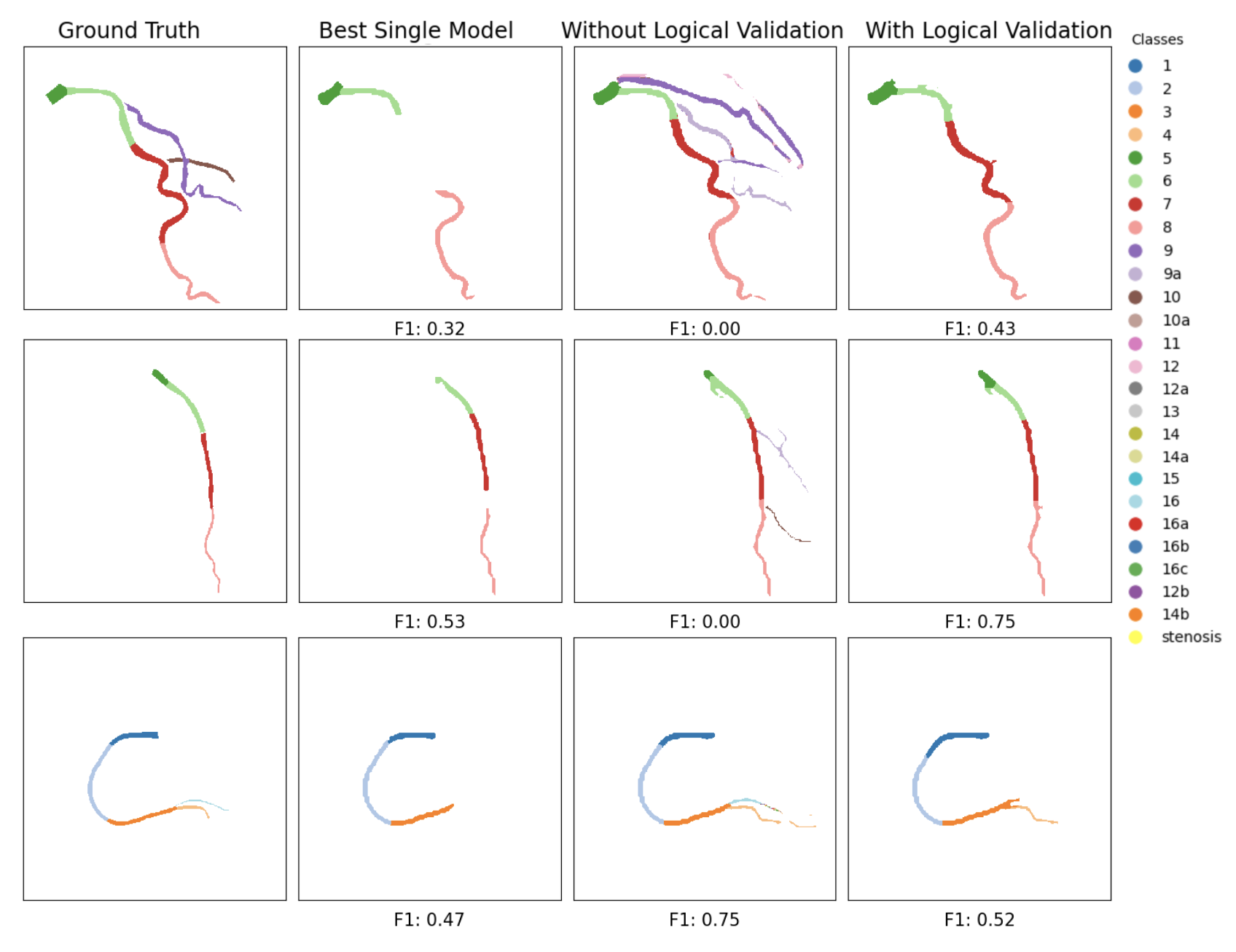}
    \caption{2D visualization of segmentation comparing best single model, with and without logical validation. For the dataset, logical validation improves the positive predictive value of segmentation.}
    \label{fig:comparison}
\end{figure}

\section{Results}
\label{Results}

\begin{figure}[ht]
    \centering
    \includegraphics[width=0.9\linewidth]{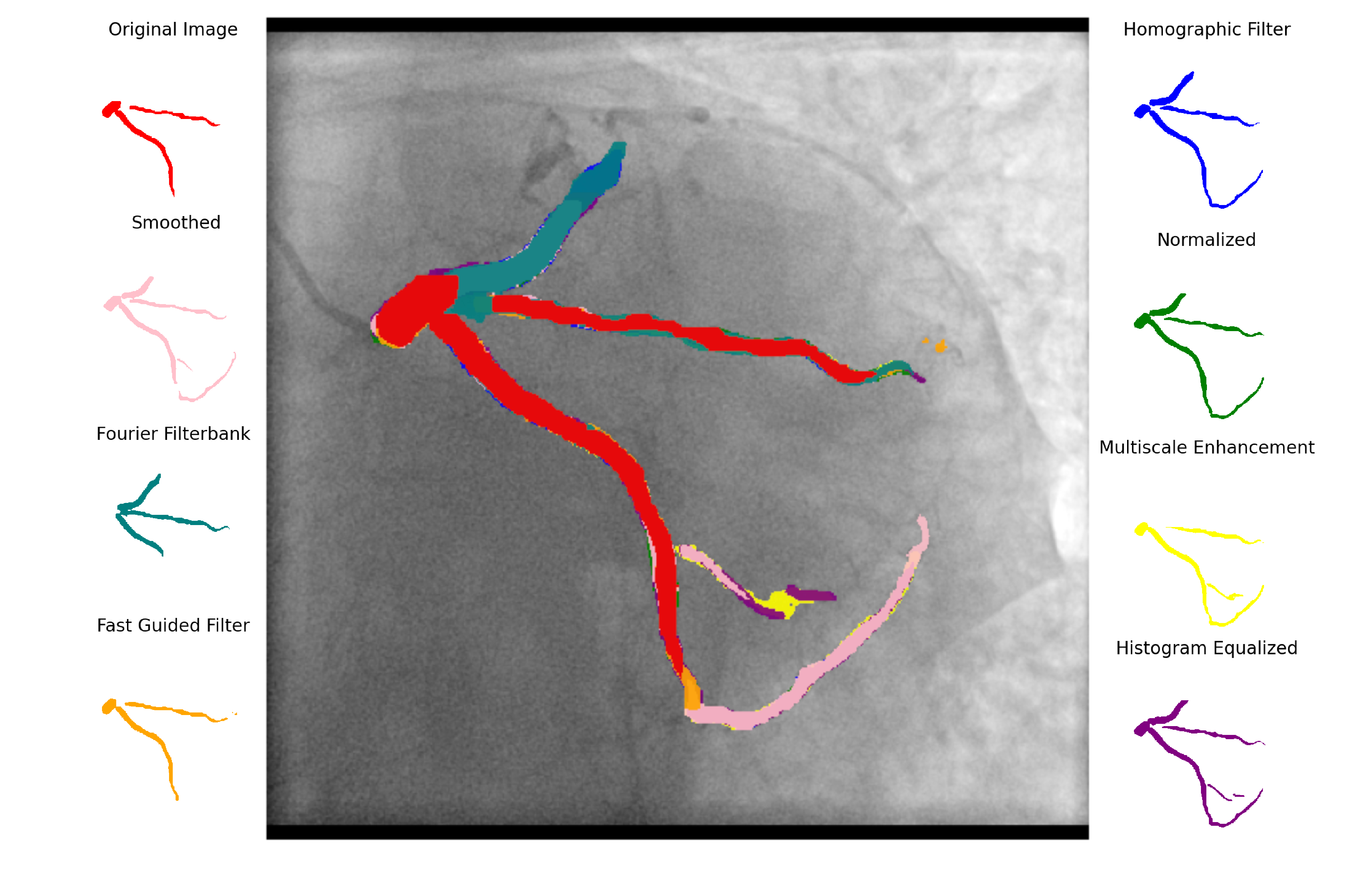}
    \caption{Application of ensemble vessel segmentation boosts detection of vessel boundaries}
    \label{figure: ensemble_mask}
\end{figure}

Table \ref{table: ablation study} presents the evaluation results achieved by the proposed segmentation method. Single model performance on original images without any pre- or post-processing produced an F1 score of 0.27. Following feature selection, the histogram equalized image \ref{figure: preprocess_example} produced the best single model performance with an F1 score of 0.31. Using ensemble prediction without logical construction described in the methods section, the best performance was 0.37. With a logically constructed arterial tree, we achieved a score of 0.411 in the publicly released test set (0.4289 in the challenge; rank 3), representing an overall 58\% improvement over the single model segmentation with the original image not using any post-processing.

\noindent
Figure \ref{fig:comparison} displays three examples of the segmentation detection results obtained by 1) best single model performance, 2) excluding the logic engine, and 3) including the logic construction. We see that logic-based, or reasoning-based anatomy validation is an important step in maximizing the potential of neural networks to correctly identify variant anatomy and drastically reduce false positives in vessel segmentation, especially when strict accuracy is required, such as in the case of this benchmark. 

\noindent \\
Next, Figure \ref{figure: ensemble_mask} illustrates the additional detection ability of subsequent preprocessing on vessel boundaries. It is noted that with the additional transformation of image features, the model has differing discrimination for the vessel boundaries and may even change the predicted class. Thus, in our current implementation, all models are used to create a vessel candidate list.

\section{Discussion}
The dataset available through the ARCADE challenge comprised static 2-D X-ray coronary angiograms. This translates as being unable to leverage sequential (frame-to-frame) information. The ARCADE dataset also had very few class instances of particular structures. Optimizing for F1 score and not clinically relevant markers is a limitation of the dataset and the challenge. However, there exists room to improve predictions for sub-segments in some cases, such as the case with segments 10, 10a, 14a, 14b, 15, 16a, 16b, and 16c for which only 25-50 training examples existed. This requires cultivating a dataset that contains more annotations of these structures.

\noindent \\
In its current state, our detection model has a tendency to overpredict vessel boundaries, which was addressed by serial erosion of the predicted segmentation. Performing this erosion as a function of image scale/field of view, however, remains to be done. We have as of this writing yet to optimize the code to reduce inference time, an important metric in implementation science for imaging study workflows. Taken together, this would facilitate a clinically viable and informative tool.

\section{Conclusion}
YOLO-Angio represents our proposed algorithm for the automated determination of coronary anatomy from single-frame X-ray coronary angiograms. We offer a novel logic engine that can improve the accuracy of other neural network models. Future work includes segmentation in time series format, which will leverage the dynamic characteristics of contrast moving through the coronary vessels. Segmentation can be combined with stenosis detection to automate the diagnosis of coronary artery disease. 

\section{Disclosures}
\label{sec:headings}

The authors report no disclosures.

\newpage
\bibliography{references}

\end{document}